\documentclass[
amsmath, %
floatfix, %
twocolumn, %
showpacs, %
reprint, %
prc, %
aps, %
nofootinbib,%
nobibnotes,%
]{revtex4-1}
\usepackage[T1]{fontenc}
\usepackage{graphicx}
\usepackage{latexsym,amsthm,dsfont,xcolor,bbm}

\usepackage[looser]{newtxtext}
\usepackage[subscriptcorrection,nosymbolsc,smallerops,bigdelims]{newtxmath}
\DeclareMathAlphabet{\mathcal}{OMS}{cmsy}{m}{n}

\usepackage{bm}

\usepackage{floatrow}
\usepackage[hyperindex,breaklinks]{hyperref}
\hypersetup{
	colorlinks,
	linkcolor={blue!100!black},
	citecolor={blue!100!black},
	urlcolor={blue!100!black}
}
\makeatletter
\renewcommand*{\eqref}[1]{%
	\hyperref[{#1}]{\textup{\tagform@{\ref*{#1}}}}%
}
\makeatother

\newcommand{\ket}[1]{\left| #1 \right>} 
\newcommand{\sket}[1]{\lvert #1 \rangle} 
\newcommand{\bket}[1]{\big\lvert #1 \big\rangle}
\newcommand{\bra}[1]{\left< #1 \right|} 
\newcommand{\sbra}[1]{\langle #1 \rvert} 
\newcommand{\bbra}[1]{\big\langle #1 \big\rvert}

\newcommand{\bmatrixel}[3]{\big\langle #1 \big\vert #2 \big\vert #3 \big\rangle}

\newcommand{\be}{\begin{equation}}
\newcommand{\ee}{\end{equation}}
\newcommand{\bea}{\begin{eqnarray}}
\newcommand{\eea}{\end{eqnarray}}
\newcommand{\bdm}{\begin{displaymath}}
\newcommand{\edm}{\end{displaymath}}

\newcommand{\MG}[1]{\color{black} #1}

\renewcommand{\mathbb}{\mathbbm}

\usepackage[sf,bf,small,raggedright]{titlesec}
\newcommand{\upemhf}{\vspace{-0.75em}}

\begin{document}
	\onecolumngrid
	{\raggedright{ \LARGE \textsf{\textbf{Spin dynamics and magneto-optical response \\[0.33em]in charge-neutral tunnel-coupled quantum dots}}}}\\[2em]
	{\raggedright{ \textsf{\textbf{ Micha{\l} Gawe{\l}czyk and Pawe{\l} Machnikowski }}}}\\[0.8em]
	{\raggedright{ \small \it Department of Theoretical Physics, Faculty of Fundamental Problems of Technology,\\Wroc\l{}aw University of Science and Technology, Wybrze\.ze Wyspia\'nskiego 27, 50-370 Wroc\l{}aw, Poland }}\\[0.8em]
	{\raggedright{ \small Email: \href{mailto:michal.gawelczyk@pwr.edu.pl}{michal.gawelczyk@pwr.edu.pl}}}\\[1em]
	{\raggedright{ \textsf{\textbf{ Abstract }}}}\\[0.5em]
	\parbox{0.8\textwidth}{ We model the electron and hole spin dynamics in an undoped double quantum dot structure, considering the carrier tunneling between quantum dots. Taking into account also the presence of an in-plane or tilted magnetic field, we provide the simulation of magneto-optical experiments, like the time resolved Kerr rotation measurement, which are performed currently on such structures to probe the temporal spin dynamics. With our model, we reproduce the experimentally observed effect of the extension of the spin polarization life time caused by the spatial charge separation, which may occur in structures of this type. Moreover, we provide a number of qualitative predictions concerning the necessary conditions for observation of this effect as well as about possible channels of its suppression, including the spin-orbit coupling, which leads to tunneling of carriers accompanied by a spin-flip. We consider also the impact of the magnetic field tilting, which results in an interesting spin polarization dynamics.}\\[1.em]

	{\noindent\rule[0.25em]{\textwidth}{1.25pt} }
	\twocolumngrid 
	
	\section{Introduction}\label{intro}
	Due to the rapid development of spintronics \cite{ZuticReview, ZuticActa} there is a need for new types of spin manipulation systems of high quality. The exploration of spin dynamics and spin coherence in semiconductors is promising thanks to the high level of the solid-state-based experimental techiques. The area of possible applications of such research is wide, including the development of new devices, like logic circuits \cite{Datta, Chiolerio}, magnetic random access memories \cite{Gallagher, Kikkawa2, Kroutvar}, semiconductor spin-based quantum computing \cite{Loss, Hawrylak}, or spin-transfer nano-oscillators \cite{Kiselev}. 
	
	The search for good structures, which could be the candidates for spin-based devices, started mostly with the investigation of the dynamics of conduction-band electrons in compound semiconductors without inversion center, for example, GaAs \cite{Wu2010}. At that time, the hole spin states did not focus so much attention mostly due to their sub-picosecond dephasing time in bulk GaAs \cite{Hilton, Shen}. However, the picture changes if one explores the spin dynamics in nanostructures, where the reduction of the system dimensionality yields the hole-spin coherence time extension, up to a few picoseconds in $p$-doped quantum wells \cite{Damen, Schneider, Lu}. This can still be enhanced by a further carrier localization occurring at low temperatures, e.g. on local interface potential fluctuations or some kind of trapping centres \cite{Saper, Korn1, Korn2, Kugler, Studer} as well as by formation of quantum dots (QDs). Additionally, hole spins are not affected by the contact-hyperfine-interaction-driven decoherence due to their $p$-like wave functions \cite{Fischer}. Finally, the high anisotropy of the hole $g$-factor in a low-dimensional GaAs, creates new possibilities of spin manipulation \cite{Machnikowski, Andlauer, Kugler2}.
	
	The possibility of the experimental investigation of spin dynamics in neutral nanostructures is limited by the exciton recombination, which, in most cases, is much faster than the actual spin dynamics to be considered \cite{Dutt2005}. For that reason, doped structures with persistently present resident electron/hole spins are commonly used and methods of resident spin polarization have been proposed \cite{Saper,Korn1,Korn2,Kugler,Studer,Kamil2}. However, the usage of doped structures comes with a number of drawbacks, first of which is the lowering of the material quality (especially optical) caused by the doping as well as problems with proper doping (particularly $p$-type) of some materials, like InGaAs \cite{Korn1}. Moreover, it has been shown that there is an intrinsic spin dephasing present in the initialization of resident spins caused both by the temporal excitation to charged excitonic complex states with subsequent recombination itself and by the phonon reservoir reaction to this carrier system evolution, which takes place during the laser pulse \cite{Gawelczyk}. This may be important in future applications but affects also the results of current experiments \cite{VarvigArXiv}, in particular those based on the resonant spin amplification effect \cite{Kikkawa, Kugler, Varwig}, where the spin coherence is crucial for the formation of the observed signal.  For that reason, the search for an undoped system with long living spins was crucial. It brought a proposal and first realizations of double quantum well/dot structures, in which, after the optical excitation, the electron-hole pair is spatially separated due to the carrier tunneling between the nanostructures enhanced by the electric field \cite{Dawson2003,Gradl2014}. This leads to the extension of carrier life time by orders of magnitude (tens of microseconds instead of less than $5~\mathrm{ns}$ for direct exciton in QWs \cite{Colocci1990}), which removes the artificial upper limit for the spin life time. 
	
	In this work we present a theoretical modeling of the spin dynamics taking place in such a system {\MG at finite temperatures}. We consider an undoped vertically stacked double quantum dot (DQD) system. The theory is also applicable to double quantum well systems, where an additional weak lateral confinement of carriers is present, as indicated by recent experimental observations \cite{Saper, Korn1, Korn2, Kugler, Studer}.
	The system is excited by a pulse of circularly polarized laser light, which generates electron-hole pairs in one of the QDs. For the simulation of the magneto-optical experiments we also take into account the presence of a magnetic field. This allows us to simulate the time resolved Kerr rotation (TRKR) signal, which is experimentally measured to probe the spin dynamics. We reproduce the effect of the extension of the spin polarization life time, which was observed experimentally \cite{Dawson2003,Gradl2014}. Then, we study the destructive impact of temperature and sample inhomogeneity on this effect. Moreover, we include the spin-orbit coupling effects, which give rise to the mixing of states with different angular momenta and, in consequence, to the probability of a spin-flip-accompanied tunneling of carriers. Finally, we also investigate the impact of the magnetic field tilting on the spin dynamics.
	
	The paper is organized as follows. First, in Sec.~\ref{model}, we provide the theoretical framework of the investigated system and derive the spin dynamics model. The results, as well as their discussion, are presented in Sec.~\ref{results}. Finally, we conclude the paper in Sec.~\ref{conclusions}. In the \hyperref[sec:appendix]{Appendix}, we present a derivation of the relative rate of tunneling with a spin flip caused by the spin-orbit interaction, which we use in Sec.~\ref{model}. 
	
	\section{The model}\label{model}
	We investigate spin dynamics in an undoped DQD system by the extension and modification of the model developed previously for trapped carriers in a single QW \cite{Machnikowski,Kugler}. The fundamental optical transition in the system consists in a generation of an electron-hole pair (exciton). We simulate the optical excitation by a pulse of circularly polarized laser light focused on one of the QDs (assumed further to be the first one). In this manner, due to the selection rules governing such pumping in this type of structures, a selected bright exciton state is pumped. 

	\begin{figure}[!tbp]
		\begin{center}
			\includegraphics[width=\columnwidth]{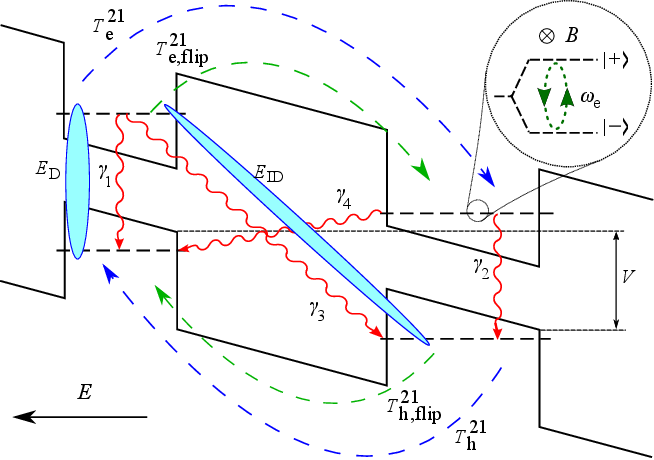}\upemhf
		\end{center}
		\caption{\label{fig:fig1}A schematic diagram of the system at $T=0~\mathrm{K}$; essential dynamical processes depicted. Dashed arrows show the tunneling process directions, including spin-preserving (blue) and spin-flip-accompanied (green) tunneling with rates equal $T^\mathrm{12/21}_\mathrm{e/h}$ and $T^\mathrm{12/21}_\mathrm{e/h,flip}$, accordingly. Red wavy arrows depict possible radiative recombination processes with rates $\gamma_{1-4}$. In the magnified part, an exemplary Zeeman splitting in the in-plane magnetic field is shown; dotted closed green loop depicts the Larmor spin precession between Zeeman eigenstates with the frequency $\omega_\mathrm{h}$. Light blue ovals represent the direct and indirect exciton binding energies for two exemplary states, $V$ is the interdot dipole energy arising from external electric field.}
	\end{figure}
	In Fig.~\ref{fig:fig1} we present a schematic energy diagram of the system and indicate the essential dynamical processes simulated in this work. The band structure is tilted due to a homogeneous electric field applied along the growth axis, which promotes the tunneling of the electron and the hole between the QDs. For simplicity the diagram corresponds to $T=0~\mathrm{K}$, at which only one way (energy preferable) tunneling for each of the carriers is possible. There are four possible radiative recombination processes in the system, each of them corresponding to the decay of one of the bright exciton states. The simulated system is placed in a homogeneous magnetic field, which causes the Larmor precession of spins. The diagram corresponds to the case of an in-plane magnetic field.
	
	The electron subsystem is described in the spin-up, spin-down basis of states (in terms of spin projection onto the axis normal to the sample plane) $\left\{ \sket{\uparrow_1}, \sket{\downarrow_1}, \sket{\uparrow_2}, \sket{\downarrow_2} \right\}$, where arrows denote the spin orientation and the number in the subscript indicates one of the two QDs. Analogously, the basis states for the holes are $\left\{  \sket{\Uparrow_1}, \sket{\Downarrow_1}, \sket{\Uparrow_2}, \sket{\Downarrow_2} \right\}$. The whole system is described in the product basis which contains the additional ground state $\sket{0}$, to allow for a description of the electron-hole pair recombination process. For convenience, we distinguish the four bright exciton states and denote them by $\sket{\mathrm{X}^\mathrm{b}_{i}}$, where $i=1, 2$ corresponds to direct excitons ($\sket{\mathrm{X}^\mathrm{b}_1}=\sket{\uparrow_1 \Downarrow_1}$, $\sket{\mathrm{X}^\mathrm{b}_2}=\sket{\uparrow_2 \Downarrow_2}$) and $i=3, 4$ to indirect ones ($\sket{\mathrm{X}^\mathrm{b}_3}=\sket{\uparrow_1 \Downarrow_2}$ and $\sket{\mathrm{X}^\mathrm{b}_4}=\sket{\uparrow_2 \Downarrow_1}$) with the spin-flipped counterpart for each of them labeled by $\mathrm{X}^\mathrm{b'}_i$. The corresponding direct and indirect dark exciton states are labeled by $\mathrm{X}^\mathrm{d}_i$ and $\mathrm{X}^\mathrm{d'}_i$, i.e. $\sket{\mathrm{X}^\mathrm{d}_1}=\sket{\uparrow_1 \Uparrow_1}$, $|\mathrm{X}^\mathrm{d'}_1\rangle=\sket{\downarrow_1 \Downarrow_1}$, \textit{etc}. 
	
	{\MG The evolution of the system, which is assumed to be in contact with a bosonic reservoir, is described in the density matrix formalism in terms of the reduced density matrix, $\rho=\mathrm{Tr}_\mathrm{R}\varrho$, where the trace is done over the degrees of freedom of the reservoir, $\varrho=\sket{\Psi}\!\sbra{\Psi}$ is the density matrix of the full system, i.e., the considered subsystem and the reservoir, and $\sket{\Psi}$ is the wave function of the full system.}
	
	We model the optical excitation by a laser Hamiltonian written in the rotating frame picture with respect to zero-field exciton energies and in the rotating wave approximation. The pump pulse is circularly polarized ($\sigma_{+}$) and energetically coupled to the first QD. The pumping laser Hamiltonian is
	\be\label{laser_hamiltonian}
	H_\mathrm{p}\left( t \right)=\frac{1}{2}f\left( t\right) e^{i\Delta t}\sket{0}\!\sbra{X^\mathrm{b}_1} + \mathrm{H.c.} \mathrm{,}
	\ee
	where $\Delta$ is the pulse detuning from the direct exciton line and $f\left( t\right)$ is the pulse envelope. One finds the density matrix of the system after the pulse in the second order with respect to the pulse amplitude 
	\begin{align}\label{ro_after_pump}
	\rho_1=&\rho_0-\frac{i}{\hbar}\int_{-\infty}^{\infty}\mathrm{d}t \left[H_\mathrm{p}\left( t \right), \rho_0 \right]\nonumber\\
	&- \frac{1}{\hbar^2}\int_{-\infty}^{\infty} \mathrm{d}t \int_{-\infty}^{\infty} \mathrm{d}t' \left[ H_\mathrm{p} \left( t \right),\left[ H_\mathrm{p}\left( t' \right), \rho_0 \right]\right]\mathrm{,}
	\end{align}
	where $\rho_0$ is the initial density matrix, assumed here to be $\rho_0=\ket{0}\!\!\bra{0}$. 
	
	{The subsequent spin dynamics is modeled using a Markovian master equation in the rotating frame picture with respect to zero-field energy gaps of QDs, but with the spin dynamics kept in the Schr{\"o}dinger picture
		\begin{align}\label{master_eq}
		\dot\rho\left( t\right)=&-\frac{i}{\hbar}\left[H_\mathrm{0},\rho\left( t\right)\right]\!+\! \sum_{i=1}^{2}\left( \mathcal{L}_{\mathrm{e}}^{i}\left[\rho\left(t\right)\right]+ 
		\mathcal{L}_{\mathrm{h}}^{i}\left[\rho\left(t\right)\right]\right)\nonumber\\
		&+ \sum_{j=1}^2 \bigg(\mathcal{L}_{\mathrm{e({\mathit{j}})}}^{t}\left[\rho\left(t\right)\right]+\sum_{\zeta}\mathcal{L}_{\mathrm{e({\mathit{j}}),flip}}^{t(\zeta)}\left[\rho\left(t\right)\right]\bigg) \nonumber\\
		&+\sum_{j=1}^2\bigg(\mathcal{L}_{\mathrm{h(\mathit{j})}}^{t}\left[\rho\left(t\right)\right]+\sum_{\zeta}\mathcal{L}_{\mathrm{h(\mathit{j}),flip}}^{t(\zeta)}\left[\rho\left(t\right)\right]\bigg)\nonumber\\
		&+ \sum_{i=1}^{4} \mathcal{L}_{\mathrm{X}_i}\left[\rho\left(t\right)\right]\mathrm{,}
		\end{align}
		where $H_0=H_\mathrm{X}+H_B$ is the Hamiltonian describing exciton energies and electron and hole spins in the magnetic field and $\mathcal{L}$s are Lindblad generators defined further. The exciton energy part is
		\be
		H_\mathrm{X}=\sum_i {E_\mathrm{X_\textit{i}}} \sum_{\xi} \bket{X^\xi_i}\!\bbra{X^\xi_i},\nonumber
		\ee
		where $\xi=\mathrm{b,~b',~d,~d'}$ and the energies, ${E_\mathrm{X_1}}=E_\mathrm{e_1}+E_\mathrm{h_1}+E_\mathrm{D}$, ${E_\mathrm{X_2}}=E_\mathrm{e_2}+E_\mathrm{h_2}+E_\mathrm{D}$, ${E_\mathrm{X_3}}=E_\mathrm{e_1}+E_\mathrm{h_2}+E_\mathrm{ID}+V$, ${E_\mathrm{X_4}}=E_\mathrm{e_2}+E_\mathrm{h_1}+E_\mathrm{ID}-V$, are composed of the single-particle electron and hole energies in QDs, $E_\mathrm{e_{1/2}}$ and $E_\mathrm{h_{1/2}}$, measured with respect to corresponding band-edges, direct and indirect exciton binding energies, $E_\mathrm{D}$ and $E_\mathrm{ID}$, and, in the case of indirect excitons, of the dipole energy $V$ arising from the external electric field (see Fig.~\ref{fig:fig1}). {\MG The Coulomb interactions between carriers as well as the presence of external electric field are effectively included in respective energy shifts. Both Coulomb interaction and electric field may also lead to the charge redistribution and consequently modify exciton radiative life time. This may be assumed to be implicitly present in the chosen radiative recombination rate, which will be introduced further in this section as a parameter of the model.}
		
		The Zeeman Hamiltonian reads
		\be\label{ }
		H_B=-\frac{1}{2} \mu_B \bm B \sum_{i=1}^{2} \left(\hat g_\mathrm{h} \bm\sigma^{i}_\mathrm{h} +g_\mathrm{e} \bm\sigma^{i}_\mathrm{e}  \right) \mathrm{,}\nonumber
		\ee
		where} $\mu_B$ is the Bohr magneton, $\hat g_\mathrm{h}$ is the hole Land{\'e} tensor with neglected in-plane anisotropy, $\hat g_\mathrm{h}=\mathrm{diag}(g_{\mathrm{h_\perp}}, g_{\mathrm{h_\perp}}, g_{\mathrm{h_\parallel}})$, where $g_{\mathrm{h_\perp}}$ and $g_{\mathrm{h_\parallel}}$ are the in-plane and axial components of $\hat g_\mathrm{h}$, respectively, $g_\mathrm{e}$ is the electron Land{\'e} factor, which is assumed to be isotropic, and $\bm\sigma^{i}_\mathrm{h}$, $\bm\sigma^{i}_\mathrm{e}$ are the vectors of Pauli matrices for the hole and electron spins in the i-th QD, respectively. We assume that the Land{\'e} factors for the hole and the electron are identical in both QDs and treat the hole as a pseudo-spin-1/2 system. For the magnetic field oriented in the $XZ$ plane, $\bm{B} = B\left(\sin\theta, 0, \cos\theta \right)$, the electron spin is simply quantized along the magnetic field direction. One also easily finds the spin quantization axis for the hole, $\hat e_\parallel=\left(\sin\phi, 0, \cos\phi \right)$, where $\tan\phi=\left({g_\perp}/{g_\parallel}\right)\tan\theta $, and the effective Land{\'e} factor for the hole is	$\tilde g_\mathrm{h} =\sqrt{(g_\mathrm{h_\perp}\sin\theta)^2 +(g_\mathrm{h_\parallel}\cos\theta)^2}$.
	
	The calculation of Zeeman eigenstates is straightforward and yields
	\begin{align}\label{}
	\ket{+_i^\mathrm{e}}&= \cos\frac{\theta}{2}\ket{\uparrow_i} +\sin\frac{\theta}{2}\ket{\downarrow_i} \mathrm{,}\nonumber \\
	\ket{-_i^\mathrm{e}}&= -\sin\frac{\theta}{2}\ket{\uparrow_i} +\cos\frac{\theta}{2}\ket{\downarrow_i} \mathrm{,}\nonumber 
	\end{align}
	for the electron subspace and analogously for holes
	\begin{align}\label{}
	\ket{+_i^\mathrm{h}\!}&= \cos\frac{\phi}{2}\ket{\Uparrow_i} +\sin\frac{\phi}{2}\ket{\Downarrow_i} \mathrm{,}\nonumber \\
	\ket{-_i^\mathrm{h}}&= -\sin\frac{\phi}{2}\ket{\Uparrow_i} +\cos\frac{\phi}{2}\ket{\Downarrow_i} \mathrm{,}\nonumber 
	\end{align}
	where the subspace is indicated in the superscript e/h, $i$ labels the QD, and $+$ and $-$ denote the upper and lower energy state, respectively.
	
	The Zeeman Hamiltonian may be rewritten using its eigenstates
	\begin{align}\label{}
	H_B = -\frac{\hbar}{2} \sum_{i=1}^{2} \big[ &\omega_{\mathrm{e}} \left(\ket{-_i^\mathrm{e}}\!\!\bra{-_i^\mathrm{e}} - \ket{+_i^\mathrm{e}}\!\!\bra{+_i^\mathrm{e}}  \right) \otimes \mathbb{1} \nonumber\\  
	 + &\omega_{\mathrm{h}} \mathbb{1} \otimes \left( \ket{-_i^\mathrm{h}}\!\!\bra{-_i^\mathrm{h}} - \ket{+_i^\mathrm{h}}\!\!\bra{+_i^\mathrm{h}}  \right) \big] \mathrm{,}\nonumber
	\end{align}
	where $\omega_{\mathrm{e}}=\mu_B B g_{\mathrm{e}}/\hbar$, $\omega_{\mathrm{h}}=\mu_B B \tilde g_{\mathrm{h}}/\hbar$ are the Larmor precession frequencies for the electron and the hole, respectively. The exciton energy Hamiltonian maintains its original diagonal form in this basis due to the spin degeneracy of excitonic energies.
	
	{\MG We neglect here the electron-hole exchange terms that would define the excitonic spectrum (fine structure) in the absence of magnetic field \cite{Bayer2002}. In a relatively strong magnetic field considered here these terms would present only a small correction to the exciton Zeeman levels.}
	
	
	The dissipative dynamics of the system, including spin relaxation, decoherence, electron and hole tunneling between the quantum QDs, and the exciton recombination, is described in the Markov limit (justified by relatively long timescales involved here) by the superoperators $\mathcal{L}$ in the universal Lindblad form \cite{Machnikowski}. The first term, $ \sum_{i=1}^{2}( \mathcal{L}_{\mathrm{e}}^{i}[\rho(t)] +\mathcal{L}_{\mathrm{h}}^{i}[\rho(t)])$ consists of the electron and hole dissipators for each QD and describes the spin relaxation and pure dephasing processes caused by the interaction between the investigated open system and its environment (e.g. phonon reservoir). The dissipators are of the form
	\begin{align}\label{ }
	\mathcal{L}_{\mathrm{\eta}}^{i}\left[\rho\right]=&~\kappa_{\mathrm{\eta}}^{i(+)} \left( \sigma_{\mathrm{\eta}}^{i(-)}\rho\sigma_{\mathrm{\eta}}^{i(+)}-\frac{1}{2}\left\{ \sigma_{\mathrm{\eta}}^{i(+)}\sigma_{\mathrm{\eta}}^{i(-)}, \rho \right\}  \right)\nonumber\\
	&+\kappa_{\mathrm{\eta}}^{i(-)} \left( \sigma_{\mathrm{\eta}}^{i(+)}\rho\sigma_{\mathrm{\eta}}^{i(-)}-\frac{1}{2}\left\{ \sigma_{\mathrm{\eta}}^{i(-)}\sigma_{\mathrm{\eta}}^{i(+)}, \rho \right\}  \right)  \nonumber \\
	&+\kappa_{\mathrm{\eta}}^{i(0)} \left( \sigma_{\mathrm{\eta}}^{i(0)}\rho\sigma_{\mathrm{\eta}}^{i(0)}-\frac{1}{2}\left\{ {\sigma_{\mathrm{\eta}}^{i(0)}}^2, \rho \right\}  \right)  \mathrm{,}\nonumber
	\end{align}
	where $\eta=\mathrm{e, h}$ denotes the carrier type, 
	\begin{align}
	\sigma_{\mathrm{e}}^{i(\pm)}&= \ket{\pm^\mathrm{e}_i}\!\!\bra{\mp^\mathrm{e}_i}\otimes\mathbb{1}  \mathrm{,}\nonumber\\
	\sigma_{\mathrm{h}}^{i(\pm)}&=\mathbb{1}\otimes \ket{\pm_i^\mathrm{h}}\!\!\bra{\mp_i^\mathrm{h}}  \mathrm{,}\nonumber\\
	\sigma_{\mathrm{e}}^{i(0)}&=\left( \ket{+_i^\mathrm{e}}\!\!\bra{+_i^\mathrm{e}}-\ket{-_i^\mathrm{e}}\!\!\bra{-_i^\mathrm{e}}  \right)\otimes\mathbb{1}  \mathrm{,}\nonumber\\
	\sigma_{\mathrm{h}}^{i(0)}&=\mathbb{1}\otimes \left( \ket{+_i^\mathrm{h}}\!\!\bra{+_i^\mathrm{h}}-\ket{-_i^\mathrm{h}}\!\!\bra{-_i^\mathrm{h}} \right)  \mathrm{}\nonumber
	\end{align}
	are the spin-flip and pure dephasing transition operators for the electron and the hole in the i-th QD, $\kappa_{\mathrm{\eta}}^{i(0)}$ the pure dephasing rates, and $\kappa_{\mathrm{\eta}}^{i(+)}$, $\kappa_{\mathrm{\eta}}^{i(-)}$ are the corresponding spin-flip rates. {\MG At zero temperature, the transitions from lower to higher states are forbidden so that $\kappa_{\mathrm{\eta}}^{i(+)}=0$. At non-zero temperatures they are} connected by the relation
	\be\label{kappa_relation}
	\kappa_{\mathrm{\eta}}^{i(+)}=\exp{\left(-\frac{\Delta E_{\eta} }{k_\mathrm{B}T} \right)}\kappa_{\mathrm{\eta}}^{i(-)} \mathrm{,}
	\ee
	which guarantees the detailed balance condition at equilibrium {\MG at temperature $T$}, with $\Delta E_{\eta}$ denoting the Zeeman energy splittings, and $k_\mathrm{B}$ the Boltzmann constant. The spin flip and dephasing rates account for any spin decoherence mechanisms present in a given system, including spin-orbit effects. {\MG Note that the spin-flip dissipators would lead to bright-dark exciton transitions in the absence of magnetic fields, while for the spin dynamics in a magnetic field, as considered here, they will induce damping of the spin precession (see Sec. 3). }
	
	The carrier tunneling is accounted for in the Markov limit by the following superoperators
		\begin{align}\label{Ltunnel}
		\mathcal{L}_{\mathrm{\eta(\mathit{j})}}^{\mathrm{t}} \left[\rho\right]=&~T_{\mathrm{\eta(\mathit{j})}}^{(12)} \left( t_{\mathrm{\eta(\mathit{j})}}^{(21)} \rho t_{\mathrm{\eta(\mathit{j})}}^{(12)}-\frac{1}{2}\left\{ t_{\mathrm{\eta(\mathit{j})}}^{(12)}t_{\mathrm{\eta(\mathit{j})}}^{(21)}, \rho \right\}  \right) \nonumber\\
		&+T_{\mathrm{\eta(\mathit{j})}}^{(21)} \left( t_{\mathrm{\eta(\mathit{j})}}^{(12)}\rho t_{\mathrm{\eta(\mathit{j})}}^{(21)}-\frac{1}{2}\left\{ t_{\mathrm{\eta(\mathit{j})}}^{(21)}t_{\mathrm{\eta(\mathit{j})}}^{(12)}, \rho \right\}  \right)\mathrm{,}  
		\end{align}
		where $\mathit{j}=\mathrm{I},\,\mathrm{II}$ denotes the position of the other carrier (which is not affected by the transition modeled here but affects the transition energy), due to the difference in the direct and indirect exciton binding energies.
		\begin{align}
		t_{\mathrm{e(I/II)}}^{(12/21)}&=\big( \sket{+^\mathrm{e}_{2/1}}\!\sbra{+^\mathrm{e}_{1/2}}+\sket{-^\mathrm{e}_{2/1}}\!\sbra{-^\mathrm{e}_{1/2}} \big)\otimes\mathbb{1}_\mathrm{I/II} \mathrm{,} \nonumber\\
		t_{\mathrm{h(I/II)}}^{(12/21)}&=\mathbb{1}_\mathrm{I/II} \otimes \big( \sket{+^\mathrm{h}_{2/1}}\!\sbra{+^\mathrm{h}_{1/2}} + \sket{-^\mathrm{h}_{2/1}}\!\sbra{-^\mathrm{h}_{1/2}}\big) \nonumber
		\end{align}
		are the tunneling transition operators, and $T_{\mathrm{\eta(\mathit{j})}}^{(12)}$, $T_{\mathrm{\eta(\mathit{j})}}^{(21)}$ are the corresponding tunneling rates in both directions for the electron and the hole, connected by the relation analogous to \eqref{kappa_relation}, with $\Delta E_{\eta(\mathit{j})}$ here equal to the energy differences between the states coupled by the tunneling process, i.e., $\Delta E_{\mathrm{e}(1)}=E_\mathrm{X_1}-E_\mathrm{X_4}$, $E_{\mathrm{e}(2)}=E_\mathrm{X_3}-E_\mathrm{X_2}$, $E_{\mathrm{h}(1)}=E_\mathrm{X_1}-E_\mathrm{X_3}$, and $\Delta E_{\mathrm{h}(2)}=E_\mathrm{X_4}-E_\mathrm{X_2}$. Finally, $\mathbb{1}_\mathrm{I/II}$ is here an identity operator acting in the subspace of the QD number of which is given in the subscript. 
	
	{\MG Spin dephasing effects have been widely studied in higher-dimensional systems, like quantum wells \cite{Nitta1997, Engels1997, Ambrosetti2009, Ambrosetti2015, Ambrosetti2016}, where spin-orbit interactions lead to polarization decay via spatial motion and momentum scattering of the carriers. For carriers fully confined in QDs, the same spin-orbit couplings induce spin relaxation by a weak admixture of states with opposite spin to the energy eigenstates \cite{Khaetskii2001}. In our model, these processes are accounted for by the spin dephasing rates $\kappa_{\mathrm{\eta}}^{i(0/+/-)}$. Additionally},	the mixing of states with different angular momenta resulting from the spin-orbit coupling leads to a nonzero probability of tunneling with a spin flip, which we take into account by addition of appropriate Lindblad generators, $\mathcal{L}_{\mathrm{\eta(\mathit{j}),flip}}^{\mathrm{t(\zeta)}}$, where $\zeta=\pm$, and the tunneling transitions are given by
		\begin{align}
		t_{\mathrm{e(\mathrm{I/II}),flip}}^{(12/21)+}&=\big( \sket{+^\mathrm{e}_{2/1}}\!\sbra{-^\mathrm{e}_{1/2}}\big)\otimes\mathbb{1}_\mathrm{I/II}\mathrm{,} \nonumber\\
		t_{\mathrm{e(\mathrm{I/II}),flip}}^{(12/21)-}&=\big( \sket{-^\mathrm{e}_{2/1}}\!\sbra{+^\mathrm{e}_{1/2}}\big)\otimes\mathbb{1}_\mathrm{I/II}\mathrm{,} \nonumber\\
		t_{\mathrm{h(\mathrm{I/II}),flip}}^{(12/21)+}&=\mathbb{1}_\mathrm{I/II}\otimes \big( \sket{+^\mathrm{h}_{2/1}}\!\sbra{-^\mathrm{h}_{1/2}}\big)\mathrm{,} \nonumber\\
		t_{\mathrm{h(\mathrm{I/II}),flip}}^{(12/21)-}&=\mathbb{1}_\mathrm{I/II}\otimes \big( \sket{-^\mathrm{h}_{2/1}}\!\sbra{+^\mathrm{h}_{1/2}}\big)\mathrm{,} \nonumber\\
		\big({t_{\mathrm{\eta(\mathrm{I/II}),flip}}^{(12/21)\pm}}\big)^\dagger&=t_{\mathrm{\eta(\mathrm{I/II}),flip}}^{(21/12)\mp},\nonumber
		\end{align}
		and their rates labeled by $T_{\mathrm{\eta(\mathrm{I/II}),flip}}^{(12/21)\pm}$. Note that tunneling processes in a given direction with opposite spin flips differ in transition energy here (by twice the Zeeman splitting) and need to be described by separate Lindblad terms.  In the \hyperref[sec:appendix]{Appendix} we derive the relative rate of the tunneling with a spin flip as compared to the spin-preserving tunneling in a typical heterostructure for the case of electron tunneling, which is given in equation \eqref{ratio}. For small self-assembled QDs this rate is negligibly small, however, in the case of weak in-plane confinement of carriers (natural QDs) at $B=5\,\mathrm{T}$ we estimate the ratio to be $T_{\mathrm{e, flip}}/T_{\mathrm{e}}\approx1/20$. We use this value in our further investigation. 
	
	The last process in the modeled dissipative dynamics is the exciton radiative recombination described by the superoperators
	\begin{align}
	\mathcal{L}_{\mathrm{X}_i}\left[\rho\right]=&~\gamma_{i} \left( \sigma_{\mathrm{X}_i} \rho {\sigma_{\mathrm{X}_i}}^\dagger-\frac{1}{2}\left\{ {\sigma_{\mathrm{X}_i}}^\dagger \sigma_{\mathrm{X}_i}, \rho \right\}  \right. \nonumber \\
	&\quad+ \left. {\sigma_{\mathrm{X}_i^{'}}} \rho  {\sigma_{{\mathrm{X}_i^{'}}}}^\dagger -\frac{1}{2}\left\{ {\sigma_{{\mathrm{X}_i^{'}}}}^\dagger {\sigma_{{\mathrm{X}_i^{'}}}}, \rho \right\}  \right)  \nonumber \\
	&+\frac{1}{2}\gamma^{(0)}_{i} \left( {\sigma^{(0)}_{\mathrm{X}_i}} \rho {\sigma^{(0)}_{\mathrm{X}_i}} -\frac{1}{2}\left\{ {{\sigma^{(0)}_{\mathrm{X}_i}}}^2, \rho \right\}  \right)  \mathrm{,}\nonumber
	\end{align}
	where $\gamma_i$ is the radiative decay rate of the $i$-th bright excitonic state, $\gamma^{(0)}_{i}$ are the additional pure dephasing rates for each exciton, and the transition operators are
	\begin{align}
	\sigma_{\mathrm{X}_i}&=\bket{0}\!\bbra{\mathrm{X}^\mathrm{b}_i} \mathrm{,}\quad \sigma_{\mathrm{X}_i'}=\bket{0}\!\bbra{\mathrm{X}^\mathrm{b'}_i} \mathrm{,}\nonumber \\
	\sigma^{(0)}_{\mathrm{X}^\mathrm{b}_i}&=\bket{\mathrm{X}^\mathrm{b}_i}\!\bbra{\mathrm{X}^\mathrm{b}_i} + \bket{\mathrm{X}^\mathrm{b'}_i}\!\bbra{\mathrm{X}^\mathrm{b'}_i}- 2 \bket{0}\!\bbra{0} \mathrm{.}\nonumber
	\end{align}
	
	To obtain the direct correspondence with experimentally measured quantities we use the density matrix in the spin-up, spin-down {\MG product} basis and construct dynamical variables, each being an average value of an appropriate operator {\MG
	\be
	A\left( t \right) = \mathrm{Tr}\big( \hat A \rho\left( t\right) \big)= \sum_{i\in \mathcal{B}_\mathrm{X}} \bmatrixel{i}{\hat A \rho \left( t\right)}{i} \mathrm{,}\nonumber
	\ee
	where the sum runs over a complete set of states. The dynamical variables of interest include} the spin polarization for electrons and holes in each QD
	\begin{align}
	\hat\Sigma^{(i)}_{\mathrm{e}}&=\frac{1}{\sqrt{2}} \sigma^{(i)}_{z} \otimes \mathbb{1}=\frac{1}{\sqrt{2}} \left( \ket{\uparrow}\!\!\bra{\uparrow} - \ket{\downarrow}\!\!\bra{\downarrow} \right) \otimes \mathbb{1} \mathrm{,}\nonumber \\
	\hat\Sigma^{(i)}_\mathrm{h}&=\frac{1}{\sqrt{2}} \mathbb{1} \otimes \sigma^{(i)}_{z}= \frac{1}{\sqrt{2}} \mathbb{1} \otimes \left( \ket{\Uparrow}\!\!\bra{\Uparrow} - \ket{\Downarrow}\!\!\bra{\Downarrow} \right) \mathrm{,} 
	\end{align}
	and the electron and hole spin coherences
	\begin{align}\label{spin_coherence}
	\hat X^{(i)}_\mathrm{e}&=\frac{1}{\sqrt{2}} \left(\ket{\uparrow}\!\!\bra{\downarrow}+ \ket{\downarrow}\!\!\bra{\uparrow} \right) \otimes \mathbb{1}  \mathrm{,} \nonumber \\
	\hat Y^{(i)}_\mathrm{e}&=\frac{-i}{\sqrt{2}} \left(\ket{\uparrow}\!\!\bra{\downarrow} - \ket{\downarrow}\!\!\bra{\uparrow}  \right) \otimes \mathbb{1} \mathrm{,} \nonumber \\
	\hat X^{(i)}_\mathrm{h}&=\frac{1}{\sqrt{2}} \mathbb{1} \otimes \left(\ket{\Uparrow}\!\!\bra{\Downarrow} + \ket{\Downarrow}\!\!\bra{\Uparrow}  \right)  \mathrm{,} \nonumber \\
	\hat Y^{(i)}_\mathrm{h}&=\frac{-i}{\sqrt{2}} \mathbb{1} \otimes \left(\ket{\Uparrow}\!\!\bra{\Downarrow} - \ket{\Downarrow}\!\!\bra{\Uparrow}  \right) \mathrm{.}  \nonumber 
	\end{align}
	
	Finally the system is probed with a linearly polarized laser pulse, which is assumed to be resonant with the fundamental transition in the probed QD. We label the time instant just before the arrival of the probe pulse by $\tau^{-}$ {\MG and the subsequent instant of measurement by $\tau$.}
	
	It is possible to collect the signal only from one selected QD, which is achieved by the tuning of the probe pulse wavelength to the transition in the selected dot, so the optical response (TRKR signal), that is, the rotation of the polarization plane of the reflected or transmitted probe pulse, is proportional to the difference between the electron and hole spin polarization \cite{Machnikowski, Yugova} in the selected QD just before the probe pulse arrival,{\MG
	\be\label{spin_polarization}
	\mathrm{TRKR}\left( \tau \right)\propto \Delta \Sigma^{(i)}\left( \tau^{-}\right)=\Delta \Sigma^{(i)}_\mathrm{e}\left( \tau^{-}\right)-\Delta \Sigma^{(i)}_\mathrm{h}\left( \tau^{-}\right) \mathrm{.}
	\ee}
	
	\section{Results and discussion}\label{results}
	The numerical solution of the equation of motion \eqref{master_eq} allows us to calculate the time dependence of spin polarization, Eq.~\eqref{spin_polarization}, in each of the QDs, which is proportional to the experimentally measured TRKR signal. In this section, we utilize this to investigate the essential features of spin dynamics in the discussed DQD system.
	
	In the first part of our simulations we set an in-plane magnetic field, $B=5~\mathrm{T}$, and effective $g$-factor values taken from experimental data \cite{Hernandez2013,Arora2013} and equal ${g}_\mathrm{e}=0.18$ and $\tilde{g}_\mathrm{e}=0.07$ for electrons and holes respectively. The electron and hole spin relaxation times for both QDs are set to $\tau_\mathrm{e}=1~\mathrm{ns}$ and $\tau_\mathrm{h}=3~\mathrm{ns}$, which are typical values for such structures \cite{Yokota2012,Roussignol1994}. The recombination times for direct and indirect excitons are set to $100~\mathrm{ps}$ and $10~\mathrm{\mu s}$ respectively. We set the direct and indirect exciton binding energies to $E_\mathrm{D}=-5\,\mathrm{meV}$ and $E_\mathrm{ID}=-1\,\mathrm{meV}$, the dipole energy $V=15\,\mathrm{meV}$, and the single-particle energy differences $\Delta_\mathrm{e}=10\,\mathrm{meV}$ and $\Delta_\mathrm{h}=1\,\mathrm{meV}$.
	
	\subsection{Tunneling time dependence}
	\begin{figure}[!tbp]
		\begin{center}
			\includegraphics[width=\columnwidth]{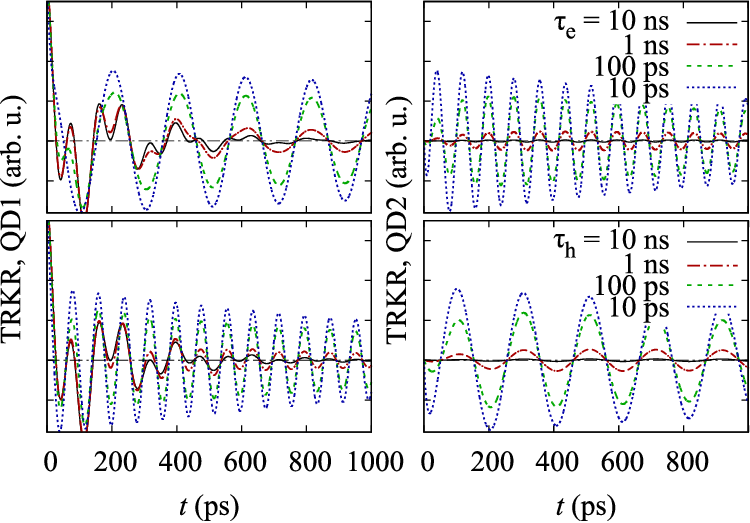}\upemhf
		\end{center}
		\caption{\label{fig:fig2}The simulated TRKR signal from the first QD (left panels) and second QD (right panels) at $1~\mathrm{K}$ under pulsed excitation of QD1 in a $5~\mathrm{T}$ in-plane magnetic field. The Land{\'e} factors are ${g}_\mathrm{e}=0.18$ and $\tilde{g}_\mathrm{h}=0.07$ for the electron and the hole, respectively. The recombination times of the direct and indirect excitons are $100~\mathrm{ps}$ and  $10~\mathrm{\mu s}$, respectively. In the upper row of panels, the case of the electron tunneling with various tunneling times is shown. The analogous plots for the hole tunneling in the bottom panels.}
	\end{figure}
	First, we investigate the extension of the spin polarization life time due to the carrier tunneling and spatial separation of the exciton. For this purpose, we simulate the TRKR signals obtained by probing both QDs under the optical pumping of the first QD at the low temperature of $1~\mathrm{K}$. In Fig.~\ref{fig:fig2}, we show the result of such simulation for various tunneling times, $\tau_\mathrm{\eta}=1/T_\mathrm{\eta}^{(12)}$, for two cases, electron or hole tunneling to the second QD, which correspond to the opposite directions of the electric field. In both cases, one can observe the effects of an interplay between the fast direct exciton recombination in the first QD and the tunneling of one of the carriers to the other QD. For slow tunneling there is no spin polarization created in the second QD, while the signal from the first QD is composed of both electron and hole spin oscillations with different periods (due to the difference in $\tilde g$ values) and it undergoes a fast exponential decay governed by the direct exciton recombination. Through intermediate situations, one arrives at the regime of dynamics dominated by the tunneling, which is dealt with when the tunneling time is comparable to or shorter than the direct exciton life time. One then observes the separation of carriers visible in the plot as single-frequency oscillations of the signals from the both QDs. The life time of the spin polarization in both QDs is extended and after some initial time (close to the tunneling time) becomes governed by the slow indirect exciton recombination. 
	
	\subsection{Temperature effects}
	\begin{figure}[!tbp]
		\begin{center}
			\includegraphics[width=\columnwidth]{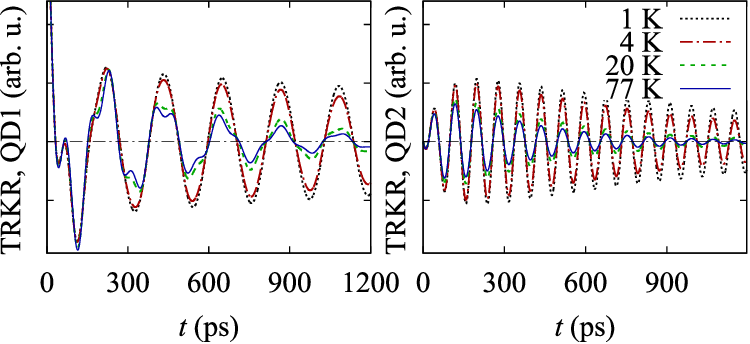}\upemhf
		\end{center}
		\caption{\label{fig:fig3}The simulated TRKR signal from the first QD (left panel) and second QD (right panel) under pulsed excitation of QD1 in $5~\mathrm{T}$ in-plane magnetic field with the electron tunneling time, $\tau_\mathrm{e}=100~\mathrm{ps}$, at various temperatures. Pure dephasing processes are absent here. Other parameters as in Fig.~\ref{fig:fig2}.}
	\end{figure}
	While in the low temperature limit discussed above one dealt effectively only with the one way carrier tunneling, from the first to the second QD, at higher temperatures the probability of the tunneling in the opposite direction becomes nonzero, which should block the desired effect of exciton separation and shorten the spin polarization life time. A simulation of this effect is shown in Fig.~\ref{fig:fig3}, where at low temperature one can see the fast electron tunneling to the second QD (a condition similar to the green dashed line plotted for $\tau_\mathrm{e}=100~\mathrm{ps}$ in Fig.~\ref{fig:fig2}). When the temperature is risen, one can observe the damping of spin polarization oscillations in the second QD. The confirmation of the fact that one really deals here with the back-tunneling can be found in the signal from the first QD, where for the time of the whole simulation a composition of two (electron and hole spin) oscillations is present. According to Eq.~\eqref{kappa_relation}, this effect becomes relevant when the thermal energy, $kT$, is comparable with the difference between the carrier energies in the two QDs. The latter, set in this simulation to $\Delta E_\mathrm{e}= 1~\mathrm{meV}$, is proportional to the applied electric field (its magnitude is of the order of $10^6~\mathrm{V/m}$ here), which up to some extent can be used to escape from this unfavorable regime.
	
	\subsection{Spin-orbit interaction effects}
	\begin{figure}[!tbp]
		\begin{center}
			\includegraphics[width=\columnwidth]{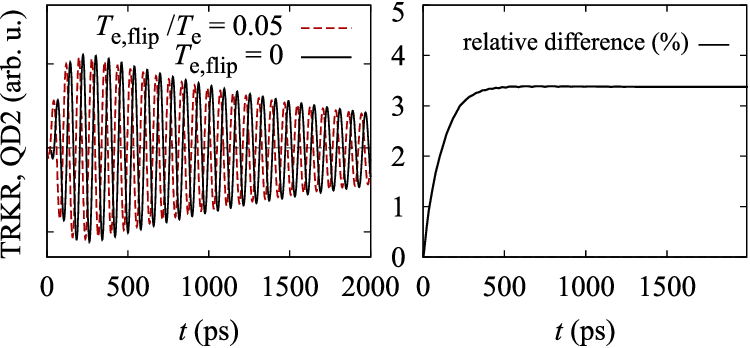}\upemhf
		\end{center}
		\caption{\label{fig:fig4}The simulated TRKR signal from the second QD (left panel) under pulsed excitation of QD1 in $1\,\mathrm{K}$. Other parameters as in Fig.~\ref{fig:fig3}. The solid black line represents the signal in the absence of the spin-orbit coupling, while the dashed red curve corresponds to the enabled spin-orbit-coupling-driven spin-flip-accompanied tunneling. One of the signals is slightly shifted in the time domain for better visibility. In the right panel, the relative difference of the two signals is shown.}
	\end{figure}
	\begin{figure}[!tbp]
		\begin{center}
			\includegraphics[width=\columnwidth]{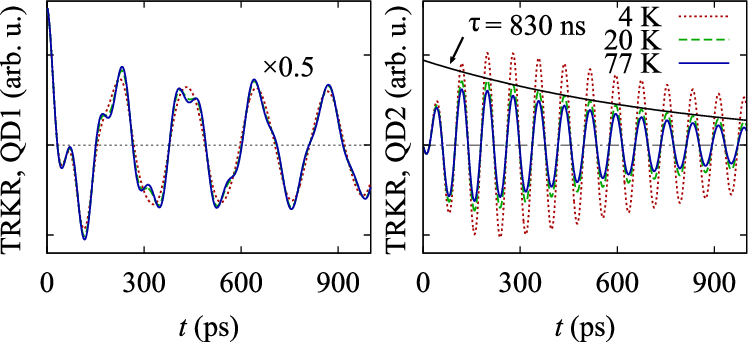}\upemhf
		\end{center}
		\caption{\label{fig:fig5}The analog of Fig.~\ref{fig:fig3} with disabled exciton recombination and enabled tunneling accompanied by a spin-flip. The additional solid black curve in the right panel is the exponential envelope of the tail of signal at $20\,\mathrm{K}$.}
	\end{figure}
	To estimate the impact of the spin-orbit interaction on the system dynamics we include the additional channel of electron tunneling accompanied by a spin flip with the rate estimated in the \hyperref[sec:appendix]{Appendix}, i.e., ${T_\mathrm{e,flip}}/{T_\mathrm{e}}={1}/{20}$. In Fig.~\ref{fig:fig4} we present the comparison of the simulated signals at $1\,\mathrm{K}$ with and without the spin-flip-accompanied tunneling (and other parameters unchanged with respect to Fig.~\ref{fig:fig3}). Since there is nearly no back-tunneling at such a low temperature, we observe here a signal loss reaching a constant value in the second QD, due to a partial loss of spin coherence at the tunneling. The amount of this loss may be found to be proportional to the ratio of tunneling rates. At higher temperatures, when a two way tunneling occurs, the process accompanied by a spin-flip leads to an accumulative signal loss, which is, however, too small to be visible as a correction to Fig.~\ref{fig:fig3}. For that reason, in Fig.~\ref{fig:fig5}, we present the results for a few values of temperature with the dominant effect of radiative recombination disabled. We estimate the lifetime of the signal at $20\,\mathrm{K}$ to be about $830\,\mathrm{ps}$, which is a small correction to the $1\,\mathrm{ns}$ electron spin lifetime put intrinsically in the model, and we find it negligible in the presence of the unavoidable recombination (compare with Fig.~\ref{fig:fig3}).
	
	\subsection{Sample inhomogeneity}
	\begin{figure}[!tbp]
		\begin{center}
			\includegraphics[width=\columnwidth]{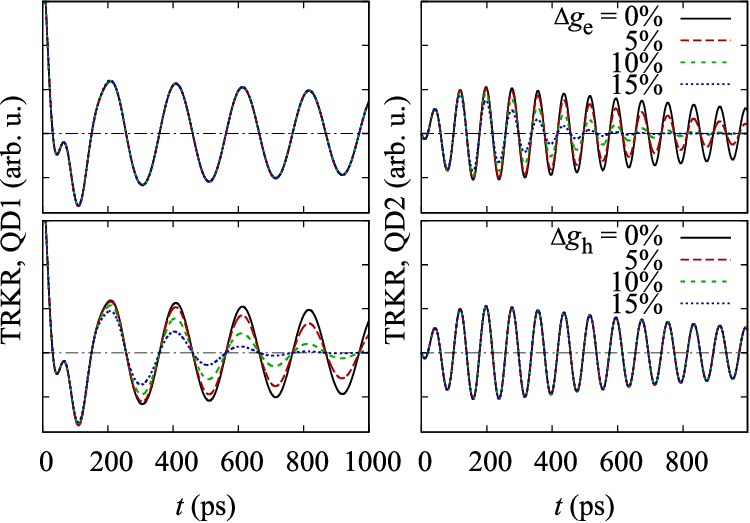}\upemhf
		\end{center}
		\caption{\label{fig:fig6}The simulated TRKR signal from the first QD (left panels) and second QD (right panels) at $1~\mathrm{K}$ under pulsed excitation of QD1 in $5~\mathrm{T}$ in-plane magnetic field. Electron tunneling time, $\tau_\mathrm{e}=100~\mathrm{ps}$. Upper panels: the simulated signal for various values of the standard deviation of the electron Land{\'e} factor from the mean value, ${g}_\mathrm{e}=0.07$. Lower panels: the analogous situation for the normal distribution of the in-plane component of the hole $g$-factor values.}
	\end{figure}
	Another disadvantageous issue, which cannot be avoided in the experiment, is the inhomogeneity of the QDs. Geometrical parameters of DQDs as well as their composition may change slightly from place to place in the ensemble on a sample, which causes, among others, the spatial fluctuations of the Land{\'e} factor values. To take this fact into account, we averaged the simulations of the TRKR signal over a normal distribution of in-plane $g$-factor component values both for holes and electrons with standard deviations $\Delta g_\mathrm{h}$ and $\Delta g_\mathrm{e}$, respectively. The result is presented in Fig.~\ref{fig:fig6}, where one can observe the impact of this inhomogeneity  separately for each type of carriers. The spin polarization oscillations are increasingly dumped with rising standard deviation of $g$. The life time of the spin polarization for both types of carriers is shortened to approximately $0.5~\mathrm{ns}$ for the standard deviation of the  Land{\'e} factor equal $\Delta g=15\%$.
	
	\subsection{Tilted magnetic field}
	\begin{figure}[!tbp]
		\begin{center}
			\includegraphics[width=\columnwidth]{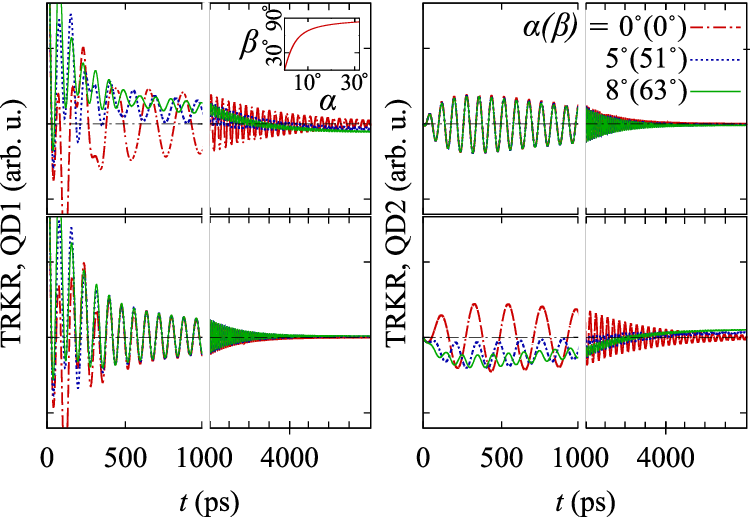}\upemhf
		\end{center}
		\caption{\label{fig:fig7}The simulated TRKR signal from the first QD (left panels) and second QD (right panels) at $1~\mathrm{K}$ under pulsed excitation of QD1 in $5~\mathrm{T}$ magnetic field tilted from the Voigt geometry by various angles, $\alpha$. In the upper panels, the case of the electron tunneling with $\tau_\mathrm{e}=500~\mathrm{ps}$ is shown; at the bottom the analogous situation for the hole tunneling. The spin pure dephasing processes are absent here.}
	\end{figure}
	Another interesting problem is the impact of the magnetic field tilting on the spin dynamics. To examine this effect, we set a constant magnetic field magnitude, $B=5~\mathrm{T}$, and vary the angle of its tilt from the Voigt geometry, $\alpha=\pi/2-\theta$. The results of this simulation are presented in Fig.~\ref{fig:fig7}. Recent experimental reports \cite{Kamil2, Soldatov2014} show that the hole $g$-tensor anisotropy  may be very high in $\mathrm{GaAs/In(Al)_{x}Ga_{1-x}As/GaAs}$ material system nanostructures, with the ratio ${g_\mathrm{h}}_\parallel/{g_\mathrm{h}}_\perp$ exceeding the factor of $10$. According to these results, in our simulation the out-of-plane and in-plane components of the hole $g$-tensor are ${g_\mathrm{h}}_\parallel=0.93$ and ${g_\mathrm{h}}_\perp=0.066$, respectively.
	
	The main feature present in the simulation for nonzero tilt angles is the deviation from the symmetric oscillations around zero polarization, which is visible in the signal from the QD1 in the case of the electron tunneling and in the signal from the QD2 in the case of the tunneling of the hole, that is in the regions dominated by the hole spin polarization. Such behavior is not observed in the signal composed mostly of the electron spin polarization. This asymmetric shape of the signal is caused by the presence of a non-precessing exponentially decaying component, which was theoretically predicted \cite{Machnikowski} and observed experimentally \cite{Kamil2} in the spin dynamics in a single doped QW. In the tilted magnetic field, the spin precession takes place around an axis, which is also tilted from the QD plane. Unlike in the exact Voigt geometry, where only the dynamics of the spin component perpendicular to the precession axis has its reflection in the simulated TRKR signal (precession damped by the spin dephasing), in a tilted field the spin component parallel to the quantization axis also has a non-zero projection on the growth axis and contributes to the signal. This makes the spin relaxation leading to thermalization between the Zeeman eigenstates (longitudinal decoherence) also visible in the simulated signal as the non-oscillating component, which contribution increases with $\alpha$.
	
	While the electron spin is quantized along the magnetic field direction, the axis of the hole spin quantization is tilted by the angle, $\beta=\pi/2-\phi$, which increases rapidly with $\alpha$ due to the strong hole $g$-tensor anisotropy (see the inset in Fig.~\ref{fig:fig7}). This explains why the non-precessing component of the spin dynamics is noticeable only in the signal composed mostly of the hole spin contribution. For $\alpha=8^{\circ}$ the tilt of the hole quantization axis reaches already $\beta=63^{\circ}$ and the signal is dominated by the nonprecessing component. In the right panels of Fig.~\ref{fig:fig7}, one can observe the dynamics of the carrier that tunnels. For the hole, it is composed of the initial increase of the signal connected with the appearance of the carrier in the QD, the non-precessing exponentially decaying component and oscillations.
	
	In a tilted field, the spin thermalization process for both types of carriers ends in the equilibrium state, which has a non-zero projection on the $z$ axis. This remaining spin polarization increases with the tilt angle of the quantization axis and is not visible for the electron in the scale of the figures. As a result one obtains a long-living (limited only by the slow indirect exciton recombination) non-oscillating hole spin polarization in one of the QDs.
	
	While the electron oscillation frequency remains unchanged upon the tilt of the magnetic field due to its isotropic $g$-factor, this is not the case for the hole, the effective Land{\'e} factor of which changes with the tilt angle. This results in the increased oscillation frequency.

	\begin{figure}[!tbp]
		\begin{center}
			\includegraphics[width=\columnwidth]{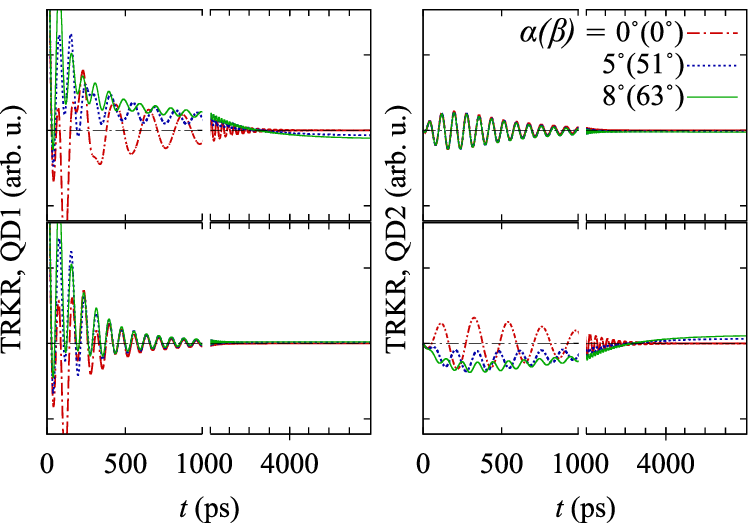}\upemhf
		\end{center}
		\caption{\label{fig:fig8}The analog of Fig.~\ref{fig:fig7} with the spin pure dephasing processes enabled; spin dephasing time equal $0.3~\mathrm{ns}$ and $1~\mathrm{ns}$ for the electron and the hole, respectively.}
	\end{figure}
	In Fig.~\ref{fig:fig8} we present an equivalent simulation, which differs only in the fact that a  fast spin dephasing process is enabled for both types of carriers with the dephasing times equal $\tau_\mathrm{e}^{(0)}=1/\kappa_\mathrm{e}^{(0)}=0.3~\mathrm{ns}$ and $\tau_\mathrm{h}^{(0)}=1~\mathrm{ns}$ for the electron and the hole, respectively. One can notice that only the oscillating component of the signal is damped here, due to the coherence decay, while the nonprecessing component behavior remains unaffected and follows the spin relaxation process, as expected.
	
	\section{Conclusions}\label{conclusions}
	We simulated the spin dynamics of optically generated carriers in the DQD system in the
	presence of magnetic field using Lindblad equations. The theoretical model developed here allows one to simulate the results of magneto-optical experiments performed on such systems, especially the signal obtained in the TRKR measurements. Our model properly reproduces the effect of the extension of the life time of the spin polarization caused by the charge separation due to the possibility of tunneling of a carrier to the other QD.  
	
	We have shown that the observed spin polarization depends essentially on the ratio of the carrier tunneling time to the direct exciton recombination time, which is a fact to be taken into account during the sample design process. Moreover, we predict that this advantageous effect can be suppressed by various factors. We show that the possibility of back-tunneling, which increases with temperature and becomes relevant when thermal energy is comparable with the difference between the energies of the direct and indirect exciton states, may totally block the considered effect. Also the impact of the ensemble inhomogeneity (QD geometry or composition) was investigated and was shown to cause the damping of the spin polarization and therefore the loss of magneto-optical signal to be observed in the experiment. In the presence of spin-orbit coupling the possibility of the tunneling accompanied by a spin-flip process leads to the damping of the signal from the target QD. However, we show that its impact is much weaker than the losses caused by the unavoidable recombination process, which makes the correction arising from the spin-orbit coupling negligible.
	
	The consideration of the magnetic field tilting showed the presence of the non-oscillating component in the TRKR signal from one of the QDs and the resulting asymmetry of spin polarization oscillations, which is considerable when the signal is composed mostly of the hole spin contribution. The hole spin relaxation in such conditions leads to a long-living non-zero signal, with the life time limited only by the slow indirect exciton recombination.
	
	The proposed model provides a complete description of the TRKR response from a DQD sample as a function of the essential parameters of the system. Hence, it can be used to extract dynamical parameters, such as carrier $g$-factors, spin life times and coherence times for investigated systems, from experimental data.
	
	\begin{acknowledgments}
	This work was funded in parts by the TEAM programme of the Foundation for Polish Science, co-financed by the European Regional Development Fund, and Polish National Science Centre Grant number 2014/13/B/ST3/04603. MG acknowledges the support from the subsidy for the Faculty of Fundamental Problems of Technology, Wroclaw University of Technology granted by the polish Ministry of Science and Higher Education to conduct research, development work and associated tasks contributing to the development of young researchers and PhD students.
		\end{acknowledgments}
	
	\renewcommand\theequation{A.\arabic{equation}}
	\section*{Appendix}\label{sec:appendix}
	In this Appendix we derive the rate of tunneling of the electron with a spin flip, which is used in Sec.~\ref{model}. Here, we perturbatively take into account the spin-orbit coupling, leading to the admixture of $p$-shell states with inverted spin to the ground state, and derive the effective phonon-mediated tunneling Hamiltonian for the case of electron tunneling. In the absence of phonons, we deal with two uncoupled subspaces, $\left\{\ket{\mathrm{D}s\!\uparrow},\ket{\mathrm{ID}s\!\uparrow},\ket{\mathrm{D}p_{-}\!\!\downarrow}, \ket{\mathrm{ID}p_{-}\!\!\downarrow} \right\}$ and $\left\{\ket{\mathrm{D}s\!\downarrow},\ket{\mathrm{ID}s\!\downarrow},\ket{\mathrm{D}p_{+}\!\!\uparrow}, \ket{\mathrm{ID}p_{+}\!\!\uparrow} \right\}$, where D and ID stand for direct and indirect exciton, respectively, $s$ and ${p_{\pm}}$ denote the orbital state of the electron with the projection of angular momentum equal, accordingly, $m=0,\pm 1$, and arrows symbolize the electron spin. We use the harmonic oscillator model for the electron states and obtain the Fock-Darwin levels $E_{\mathrm{D}s\uparrow}$, $E_{\mathrm{ID}s\uparrow}=E_{\mathrm{D}s\uparrow}-\Delta$, $E_{\mathrm{D}p_{-}\!\uparrow}=E_{\mathrm{D}s\uparrow}+\omega_{sp_{-}}$, $E_{\mathrm{ID}p_{-}\!\uparrow}=E_{\mathrm{D}s\uparrow}+\omega'_{sp_{-}}-\Delta$, for the first subspace. In the second subspace, $E_{\mathrm{D}s\downarrow}=E_{\mathrm{D}s\uparrow}-g\mu_\mathrm{B}B$ and the other relations are analogous. Here, $\omega_{sp_{\pm}}=\hbar\omega_\mathrm{B}\mp\left(\hbar\omega_\mathrm{c}+g\mu_\mathrm{B}B\right)$, $\omega_\mathrm{B}=({\omega_0^2+\frac{1}{4}\omega_\mathrm{c}^2})^{1/2}$, $\omega_\mathrm{c}={eB}/{m^{*}}$, and $\omega_0$ is the zero-field oscillator frequency.
	
	As physics in both subspaces is similar, we only need to investigate one of them by writing the tunneling Hamiltonian
	\be
	H_T=-\sum_{\alpha=s,p_{-}} \sum_{\sigma} t_\alpha \ket{\mathrm{D}\alpha\sigma}\!\!\bra{\mathrm{ID}\alpha\sigma} + \mathrm{H.c.}\nonumber,
	\ee
	where $t_\alpha>0$ are the tunneling couplings and there is no coupling between $s$ and $p$ states due to the difference in their symmetry. The spin-orbit Hamiltonian leading to the $s$-$p$ mixing, which is dominant among other spin-orbital spin-flip mechanisms \cite{Khaetskii2001}, reads
	\be
	H_\mathrm{so}=\xi\left(\ket{\mathrm{D}s\!\uparrow}\!\!\bra{\mathrm{D}p_{-}\!\!\downarrow}+ \ket{\mathrm{ID}s\!\uparrow}\!\!\bra{\mathrm{ID}p_{-}\!\!\downarrow}\right) + \mathrm{H.c.},\nonumber
	\ee
	with the coupling $\xi=-i\beta p_0$, where $\beta={\gamma}\left<k_z^2\right>/{\hbar}\approx 10\,{\mathrm{nm}}/{\mathrm{ps}}$ for typical GaAs heterostructures \cite{Khaetskii2001}, $p_0={\hbar}/{l_0}$, $l_0=\sqrt{{\hbar}/{m^{*}\omega_\mathrm{B}}}$, and we assume the coupling to be equal in both QDs. As the states are strongly localized along the $z$ axis, we may neglect the spin-orbit couplings between direct and indirect states.
	
	Finally, the Hamiltonian is $H=H_\mathrm{T}+H_\mathrm{so}$ and we assume $\Delta\sim t_\alpha$; $\omega_{sp_{-}},\omega'_{sp_{-}}\gg\Delta,\xi$. Treating $H_\mathrm{so}$ as a small correction in a quasi-degenerate perturbation theory \cite{Cohen1998AtomPhoton}, we find the perturbed Hamiltonian
	\be\label{transform}
	H'=THT^\dagger;~~T=e^{iS};~~S=\lambda S_1+\lambda^2 S_2+...,
	\ee
	where $\lambda$ is a formal expansion parameter. In the first order, the only nonzero matrix elements are
	\begin{align}
	\bra{ls\!\uparrow}i\lambda S_1\ket{lp_{-}\!\!\downarrow}&=\frac{\bra{ls\!\uparrow}i\lambda H_\mathrm{so}\ket{lp_{-}\!\!\downarrow}}{-\omega^{(l)}_{sp_{-}}}=-\lambda\frac{\xi}{\omega^{(l)}_{sp_{-}}},\nonumber\nonumber
	\end{align}
	and their Hermitian conjugates, where $l=\mathrm{D,ID}$, and further we assume $\omega^{(l)}_{sp_{-}}$ to be equal. Analogous formula holds in the second subspace.
	
	The phonon bath is taken into account by the free phonon Hamiltonian, $H_\mathrm{ph}=\sum_{\bm{k},s}\hbar\omega_{\bm{k},s}b^\dagger_{\bm{k},s}b_{\bm{k},s}$, where $s$ denotes a phonon branch, and by the exciton-phonon coupling
	\be
	H_\mathrm{X\mbox{-}ph}=\sum_{\bm{k},s} \sum_{l=\mathrm{D,ID}} \sum_{{\alpha,\beta=s,p_\pm}} \!\!\!\sigma_{l\alpha\beta} F_{l\alpha\beta}\left(\bm{k},s\right)\left(b_{\bm{k},s}+b^\dagger_{-\bm{k},s}\right), \nonumber
	\ee
	where $\sigma_{l\alpha\beta}=\sigma^\dagger_{l\beta\alpha}=\ket{l\alpha}\!\!\bra{l\beta}$, $F_{l\beta\alpha}\left(\bm{k},s\right)=F^{*}_{l\alpha\beta}\left(-\bm{k},s\right)$ is the coupling constant, and due to the carrier localization there are no D-ID transitions. This Hamiltonian, after transformation \eqref{transform} keeps its original form with the substitution $\sigma_{l\alpha\beta}\rightarrow\sigma'_{l\alpha\beta}=T\sigma_{l\alpha\beta}T^\dagger$, which in the first order is $\sigma'_{l\alpha\beta}\approx \sigma_{l\alpha\beta}+\left[iS_1,\sigma_{l\alpha\beta}\right]$, and we find the spin-orbit-induced correction to be
	\begin{multline}
	\tilde{H'}_\mathrm{X\mbox{-}ph}\approx\sum_{\bm{k}s} \sum_{l=\mathrm{D,ID}} \xi\left[\frac{F_{lsp_{+}}\!\left(\bm{k},s\right)}{\omega_{sp_{+}}}-\frac{F_{lp_{-}s}\!\left(\bm{k},s\right)}{\omega_{sp_{-}}}\right]\nonumber\\
	\times\ket{ls\!\uparrow}\!\!\bra{ls\!\downarrow}\left(b_{\bm{k}s}+b^\dagger_{-\bm{k}s}\right)+\mathrm{H.c.}\nonumber
	\end{multline}
	We assume here $F_{lsp_{+}}\approx F_{lp_{-}s}$ (which holds exactly for Fock-Darwin states) and use an approximation ${\omega^{-1}_{sp_{-}}} - {\omega^{-1}_{sp_{+}}}\approx {2}{(\hbar\omega^{(l)}_\mathrm{B})^{-2}}(\frac{\hbar e}{m^{*}} +g\mu_\mathrm{B}) B$, 
	where $\omega^{(l)}_\mathrm{B}$, corresponding to the electron in one of the QDs, are further assumed to be equal due to similarity of QDs, and the exciton-phonon coupling constants to differ only by a factor arising from the translation, i.e., $F_{lsp_{+}}\!\left(\bm{k},s\right)\equiv e^{ik_z z_l} F_1\left(\bm{k},s\right)$, where $z_\mathrm{D}=-d/2$, $z_\mathrm{ID}=d/2$, and $d$ denotes the interdot distance. Next, we diagonalize the tunnel coupling by defining the $s$-shell basis, $\ket{+\sigma} = \cos{\frac{\theta}{2}} \ket{\mathrm{D}s\sigma} + \sin{\frac{\theta}{2}} \ket{\mathrm{ID}s\sigma}$, $\ket{-\sigma} = -\sin{\frac{\theta}{2}} \ket{\mathrm{D}s\sigma} + \cos{\frac{\theta}{2}} \ket{\mathrm{ID}s\sigma}$, where $\tan{\theta}=-{2t_s}/{\Delta}$ and $E_{\pm}=\pm\sqrt{t^2_s+({\Delta}/{2})^2}-{\Delta}/{2}$. The tunneling transitions accompanied by a spin flip are generated by the off-diagonal part of Hamiltonian, which now reads
	\begin{align}
	\tilde{H'}^\mathrm{off\mbox{-}diag}_\mathrm{X-ph}=&~\sum_{\bm{k}s} \frac{2\beta p_0}{\hbar^2 \omega^2_\mathrm{B}}\left(\frac{\hbar e}{m^{*}}+g\mu_\mathrm{B} \right)B\sin{\theta}\sin{\frac{k_z d}{2}}\nonumber \\
	&\times F_1\left(\bm{k},s\right) \left(\ket{+\!\uparrow}\!\!\bra{-\!\downarrow} + \ket{-\!\uparrow}\!\!\bra{+\!\downarrow}\right) \nonumber \\
	&\times \left(b_{\bm{k},s}+b^\dagger_{-\bm{k},s}\right) + \mathrm{H.c}.\nonumber
	\end{align}
	and the Fermi golden rule leads to the transition rate
	\begin{align}\label{rate1}
	T_\mathrm{e,flip}=&~\frac{2\pi}{\hbar} \left[ \frac{2\beta p_0 \sin{\theta} B}{\hbar^2 \omega^2_\mathrm{B}} \left(\frac{\hbar e}{m^{*}} +g\mu_\mathrm{B}\right)\right]^2\nonumber\\
	&\times \sum_{\bm{k}s} \left|F_1\left(\bm{k},s\right) \right|^2 {\sin^2{\frac{k_z d}{2}}} \delta\!\left(\hbar \omega_{\bm{k}s}-\Delta E\right)\nonumber\\
	&\times \left[n_\mathrm{B}\left(\frac{\Delta E}{\hbar}\right)+1\right],
	\end{align}
	where $\Delta E=E_{+}-E_{-}$.
	
	The tunneling without spin flip is generated by the Hamiltonian
	\be
	H^{(0)}_\mathrm{X-ph}=\sum_{\sigma l} \sum_{\bm{k}s} F_{lss}\left(\bm{k},s\right) \ket{ls\sigma} \!\! \bra{ls\sigma} \left(b_{\bm{k},s}+b^\dagger_{-\bm{k},s}\right)\nonumber,
	\ee
	the off-diagonal part of which, after a treatment analogous  as above, leads to the following transition rate
	\begin{align}\label{rate2}
	T_\mathrm{e}=&~\frac{2\pi}{\hbar} {\sin^2{\theta}} \sum_{\bm{k}s} \left|F_0\left(\bm{k},s\right) \right|^2 {\sin^2{\frac{k_z d}{2}}}\nonumber \\
	&\times\delta\!\left(\hbar \omega_{\bm{k}s}-\Delta E\right) \left[n_\mathrm{B}\left(\frac{\Delta E}{\hbar}\right)+1\right],
	\end{align}
	where $F_{lss}\!\left(\bm{k},s\right)\equiv e^{ik_z z_l} F_0\left(\bm{k},s\right)$.
	
	Finally, assuming that sums in \eqref{rate1} and \eqref{rate2} are of the same order of magnitude, we arrive at the ratio of tunneling rates equal
	\be\label{ratio}
	\frac{T_\mathrm{e,flip}}{T_\mathrm{e}}=\left[\frac{2\beta p_0}{\hbar^2 \omega^2_\mathrm{B}} \left(\frac{\hbar e}{m^{*}} +g\mu_\mathrm{B}\right) B \right]^2.
	\ee
	At  $B=5\,\mathrm{T}$, for small QDs, for which $\hbar\omega_\mathrm{B}\sim 50\,\mathrm{meV}$, $\beta p_0=\hbar\beta/l_0\sim 2\,\mathrm{meV}$, $g\mu_\mathrm{B}\ll{\hbar e}/{m^{*}}$, and ${\hbar e B}/{m^{*}}\sim 5\,\mathrm{meV}$ the ratio is ${T_\mathrm{e,flip}}/{T_\mathrm{e}}\sim8\cdot 10^{-5}$. In the case of weak localization (natural QDs), we estimate $\hbar\omega_\mathrm{B}\sim 5\,\mathrm{meV}$, and $\beta p_0=\hbar\beta/l_0\sim 0.5\,\mathrm{meV}$, which leads to the ratio of tunneling rates on the order of ${T_\mathrm{e,flip}}/{T_\mathrm{e}}\sim 1/20$. 
	\setcounter{section}{1}
	
	\providecommand{\newblock}{}

\end{document}